\documentclass{elsart}
\usepackage{amssymb}   
\usepackage{graphicx}
\begin{document}
\begin{frontmatter}

\title{TeV photons and Neutrinos from giant soft-gamma repeaters flares}

\author[label1]{Francis Halzen}, 
\ead{halzen@pheno.physics.wisc.edu}
\author[label1]{Hagar Landsman}, 
\ead{hagar.landsman@icecube.wisc.edu}
\author[label1,label4]{Teresa Montaruli\corauthref{cor1}}
\address[label1]{University of Wisconsin, Chamberlin Hall, Madison, WI 53706}
\address[label4]{On leave of absence from 
Universit\'a di Bari and INFN}
\ead{tmontaruli@icecube.wisc.edu}
\corauth[cor1]{Corresponding author.}

\begin{abstract}

During the last 35 years three giant flares were observed from so-called 
Soft Gamma Repeaters (SGR's). They are assumed to be associated with star-quakes 
of pulsars accelerating electrons and, possibly, protons to high energy in the 
huge magnetic fields as inferred from the observations. 
Because of this and the observation of non-thermal emission it has been 
speculated that they may be cosmic ray accelerators producing gamma-rays 
up to TeV energies. 
Neutrino telescopes, such as AMANDA and
the ANTARES now under construction,
could be used as TeV-$\gamma$ detectors for very short emissions 
by measuring underground muons produced in $\gamma$ showers. 
We estimate signal and background rates for TeV photons from SGR giant flares 
in AMANDA, and we provide an estimate of the gamma shower events
that Milagro could detect.

Moreover, we consider that, if hadrons are accelerated in these
sources, high energy neutrinos would be produced together with photons.
These may be detected in neutrino telescopes using neutrino-induced cascades
and upgoing muons.

We argue that the Antarctic 
Muon and Neutrino Detector Array AMANDA may have observed the December 27, 2004 
giant flare from the soft gamma-ray repeater SGR 1806-20 if 
the non-thermal component of the spectrum extends to TeV energies 
(at present the actual data is subject to blind analysis). 
Rates should be scaled by about two orders of magnitude in
km$^3$ detectors, such as IceCube, making SGR flares sources of primary interest.
\end{abstract}
\end{frontmatter}

\section{Introduction}
\label{Sec1}

Soft gamma repeaters are X-ray pulsars that sporadically emit energies 
typically of the order of $10^{41} D_{10}^2$ ergs in X-rays and soft 
$\gamma$-rays, with the distance D in units of 10 kpc, in short bursts 
lasting hundreds of milliseconds \cite{review,review1}. 
Their steady emission is periodic with typical periods of 5-10 seconds 
and luminosities between $10^{35} \div 10^{36}$ ergs/s. 
These objects are thought to be magnetically-powered spinning neutron stars 
also known as magnetars.

The first SGR was observed on Jan. 7, 1979 when
a burst of soft gamma-rays, lasting about 0.25 s, was detected by the 
Venera space-craft from SGR 1806-20 \cite{review}. To date
five SGR's have been identified.
The giant bursts are thought to result from the rearrangement 
of the huge magnetic field inferred to be of the order of $B \sim 10^{15}$ Gauss 
from their spin down rates  \cite{Kouveliotou}. Star-quakes are thought 
to fracture the rigid crust and cause these outbursts. 
The flares result from the formation and dissipation of strong localized 
currents resulting from magnetic field rearrangements associated with the 
quakes.

SGR's have been classified as separate phenomena from gamma-ray bursts (GRB's) 
and anomalous X-ray pulsars (AXP's) \cite{Kaspi} despite similarities. 
SGR bursts are less frequent than GRB's and, unlike GRB's, they do repeat and 
show clear periodicity. It is not excluded however that at least
some of the short GRB's are extra-galactic SGR's and, vice-versa, that some 
of the intense flares of SGR's are galactic GRB's. SGR's are very similar 
to the 8 observed AXP's, the main difference being the lack of bursts for 
the latter. Recently, though, SGR-like bursts were 
observed from at least one, and possibly two AXP's \cite{Kaspi}. 
Hence SGR's and AXP's might belong to the same class of neutron stars.

Like GRB's, SGR's also produce non-thermal radiation during the short bursts. 
Because of this and the large luminosities produced, it has been speculated 
that these objects accelerate cosmic rays to high energy. The sudden changes 
of the large magnetic fields would accelerate protons or nuclei that produce 
neutral and charged pions in interactions with thermal radiation. 
These would be the parents of gamma-rays and neutrinos reaching high energies.

We here propose that these high energy gamma-rays can be detected by 
underground neutrino telescopes and by experiments, such as Milagro 
\cite{Milagro} and the Tibet array \cite{Tibet}. 
(The premier instruments for TeV gamma-ray detection, 
air Cherenkov telescopes, must point at an SGR in order to detect it and 
this is unlikely for such a serendipitous event.) Underground detectors 
have the capability to detect high energy muons produced in air showers 
initiated by TeV-energy gamma-rays at the top of the atmosphere. 
Though gamma-ray showers are muon-poor, the large number of secondary 
photons produced will occasionally produce pions rather than 
electron-positron pairs. These decay into muons that can reach 
large underground detectors such as Baikal \cite{Baikal}, AMANDA/IceCube
\cite{AMANDA,IceCube} and future Mediterranean detectors 
\cite{ANTARES,NEMO,km3,NESTOR}.  
 
TeV gamma-rays could be produced in electromagnetic process
and also in hadronic interactions on the thermal radiation, and in this
case they would be accompanied by a flux of neutrinos. A model for 
neutrino production from magnetars in their steady phase of periodic 
emission has been proposed \cite{Zhang}. Here we concentrate instead on 
the possibility that short bursts of neutrinos are produced during the rare 
violent giant flares. Because of the negligible background during the short 
time and in the direction of the SGR, the instruments have a much larger 
sensitivity for such events. 

This paper is organized as follows: 
in Sec.~\ref{Sec2} we review the photon observations of the giant flares. 
In Sec.~\ref{Sec3} we argue that they should produce detectable events in 
earth-based experiments provided their
spectra extend to TeV energies not accessible to satellites. 
For illustration we calculate the rates of muons from gamma
showers on Milagro surface \cite{Milagro}
and the number of muons from gamma showers reaching the 
AMANDA detector \cite{AMANDA}. In Sec.~\ref{sec:neu} we estimate the number of 
upward-going muons and showers produced by neutrinos in the 
same detector. 
Similar rates are expected in neutrino telescopes of similar depth 
such as ANTARES \cite{ANTARES}.

\section{Giant bursts from Soft Gamma Repeaters: Observations}
\label{Sec2}

In recent years SGR giant bursts have been observed from 3 objects: 
SGR 0526-66 on May 5, 1979, SGR 1900+14 on August 27, 1998 and 
SGR 1806-20 on December 27, 2004 \cite{review}. Their UT
times and equatorial coordinates are given in Tab.~\ref{tab1}.

\begin{table}[htb]
\begin{tabular}{|l|c|c|c|l|}
\hline
{\bf {\small Source}} & {\bf {\small RA}} & {\bf {\small Declination}} & 
{\bf {\small Distance}} & {\bf {\small UT}}\\ 
& {\bf \small (J2000)} & {\bf \small (J2000)} & 
{\bf \small (kpc)} & \\ \hline
{\small SGR 0526-66} & {\small $05^{h} 26^{m} 00.89^{s}$} & 
{\small $-66^{\circ} 04^{'} 36.3^{''}$} & {\small 50} & {\small May 5, 79 15:51}\\
{\small SGR 1900+14} & {\small $19^{h} 07^{m} 14.24^{s}$} & 
{\small $+09^{\circ} 19^{'} 20.1^{''}$} &{\small 14.5} & {\footnotesize Aug. 27, 98 10:22:15.7} \\
{\small SGR 1806-20} & {\small $18^{h} 08^{m} 39.32^{s}$} & 
{\small $-20^{\circ} 24^{'} 39.5^{''}$} & {\small 15.1} & {\footnotesize Dec. 27, 04 21:30:26.65} \\
\hline
\end{tabular}
\caption{\label{tab1} 
Coordinates and UT times of the giant flares \protect\cite{review}.
UT time for SGR 1900+14 is from \protect\cite{Feroci} and for 
SGR 1806-20 from \protect\cite{GCN2920}. 
Positions (right ascension RA and declination) 
and distances are from \protect\cite{Vrba} for SGR 1900+14 and 
from \protect\cite{Kaplan} and \protect\cite{Corbel} for SGR 1806-20, 
respectively.}
\end{table} 

Giant flares are distinguished from common SGR bursts by their extreme 
energies ($\sim 10^{44}$ ergs/s) emitted during the first second followed by subsequent 
emission lasting several minutes showing pulsations associated with the  
neutron star. While the total energy emitted is 6-7 orders of magnitude 
less intense than for GRB's, their distance 
is typically a factor of $10^5$ smaller resulting in a flux increased 
by a factor $10^{10}$ relative to cosmological GRB's. 
The opportunity for a detection is obvious.

The bulk emission of SGR's can be routinely explained by Optically 
Thin Thermal Bremsstrahlung (OTTB). We focus on the non-thermal component 
observed in giant flares with a hard power law (PL) spectrum, 
$\frac{dN}{dE} \propto E^{-1.5}$ \cite{review,Feroci}. This is harder 
than the Band spectrum observed for a typical GRB \cite{Band}. 
Because of the energetics and the non-thermal spectra giant flares should 
be prime suspects for accelerating cosmic rays and for producing associated 
fluxes of photons and neutrinos of TeV-energy, possibly higher.

An interesting feature is that both SGR 1900+14 and SGR 1806-20 are 
associated with clusters of massive stars \cite{review,Vrba,Corbel}, 
while the other giant flare emitter, SGR 0526-66, 
is associated with the supernova remnant N49 in the Large Magellanic Cloud. 
It has been suggested that association of massive stars, such as Cygnus OB2, 
can host highly magnetized pulsars that accelerate heavy nuclei up
to high energies \cite{Bednarek}.
TeV $\gamma$'s and $\nu$'s can be produced by cosmic rays by interacting 
with the innermost parts of the winds of massive O and B stars \cite{Torres}. 
Moreover, relativistic nuclei photo-disintegrate in the intense radiation 
fields producing neutrinos from the decay of neutrons \cite{Anchordoqui}. 
For a review of galactic neutron stars in high density environments: see \cite{Burgio}.
In a recent paper \cite{Piran} it has also been suggested that the 
December 27, 2004 flare has similar features to a GRB and can be similarly 
explained as a relativistically expanding fireball with baryon loading. 
Proton acceleration will inevitably lead to neutrino emissions.

The December 27, 2004 flare exceed previous flares in intensity by over 
one order of magnitude. It was initially detected by INTEGRAL \cite{GCN2920}, 
then by Konus-Wind \cite{GCN2922} and later by RHESSI \cite{GCN2936}. 
While the germanium detectors operating from 3 keV to 15 MeV
were saturated, RHESSI measured the flux using the particle detector with a 
time resolution of 0.125~s in two energy channels with thresholds of 
65~keV and 650~keV. 
These data indicate significant emission above 650~keV for $\sim 0.25$~s 
during the giant flare. A conservative lower limit on the total fluence of 
the primary giant peak is 0.1 ergs/cm$^{2}$, more than 
one order of magnitude larger 
than reported for SGR 1900+14 ($5.5 \cdot 10^{-3}$ ergs/cm$^{2}$ in the 
first $\sim 0.35$~s \cite{Mazets}). Assuming a distance of $\sim 15$~kpc 
for SGR 1806-20 \cite{Corbel} and isotropic emission, a lower limit on 
the total hard X-ray/gamma-ray 
energy released is $8 \cdot 10^{45}$ ergs during the giant flare.

So far no spectral analysis has been reported for the 27 Dec., 2004 
giant flare during which the AMANDA neutrino telescope was taking data. 
In Sec.~\ref{sub1} we study instead the features of the giant flare 
SGR 1900+14 on August 27, 1998 during which the AMANDA detector had been 
partially completed and and was taking data with 10 strings; we refer to it 
as AMANDA~B-10.

\subsection{The features of the SGR 1900+14 giant flare}
\label{sub1}

SGR 1900+14 was first observed in Jan. 1979 in the Aquila constellation. 
Thought initially to be located near SNR G42.8+0.6 at $\sim 5$ kpc, 
it is now believed to be associated with a massive star cluster at 
14.5 kpc \cite{Vrba}. The position of the persistent emitter is given 
in Tab.~\ref{tab1} \cite{Vrba}. Its period of steady X-ray emission is 
5.16 s with a luminosity of the order of $2-3.5 \cdot 10^{35}$ ergs/s. 
The quiescent energy spectrum can be fitted by a Black Body (BB) function 
with a temperature of 0.43 keV and a Power Law (PL) shape with a negative 
spectral index in the range $-1.0 \div -2.5$ \cite{review}. 
Various flares with peak luminosities between $10^{38} \div 10^{41}$ ergs/s 
and durations of the order of 0.1 s have been observed. 
SGR 1900+14 emitted a giant flare on Aug. 27, 1998 
\cite{Feroci,Mazets,Feroci1,Guidorzi}. Two relatively high fluence events 
occurred on Aug. 29, 1998 \cite{Ibrahim} and on 
Apr. 28, 2001 \cite{Lenters}. For reasons already explained in 
Sec.~\ref{Sec2}, we focus on the giant flare whose spectral properties 
have been well studied, including those of the hard non- thermal component 
\cite{Feroci,Guidorzi}.

\begin{figure}[htb]
\begin{tabular}{cc}
\includegraphics[width=7.5cm,height=8.cm]{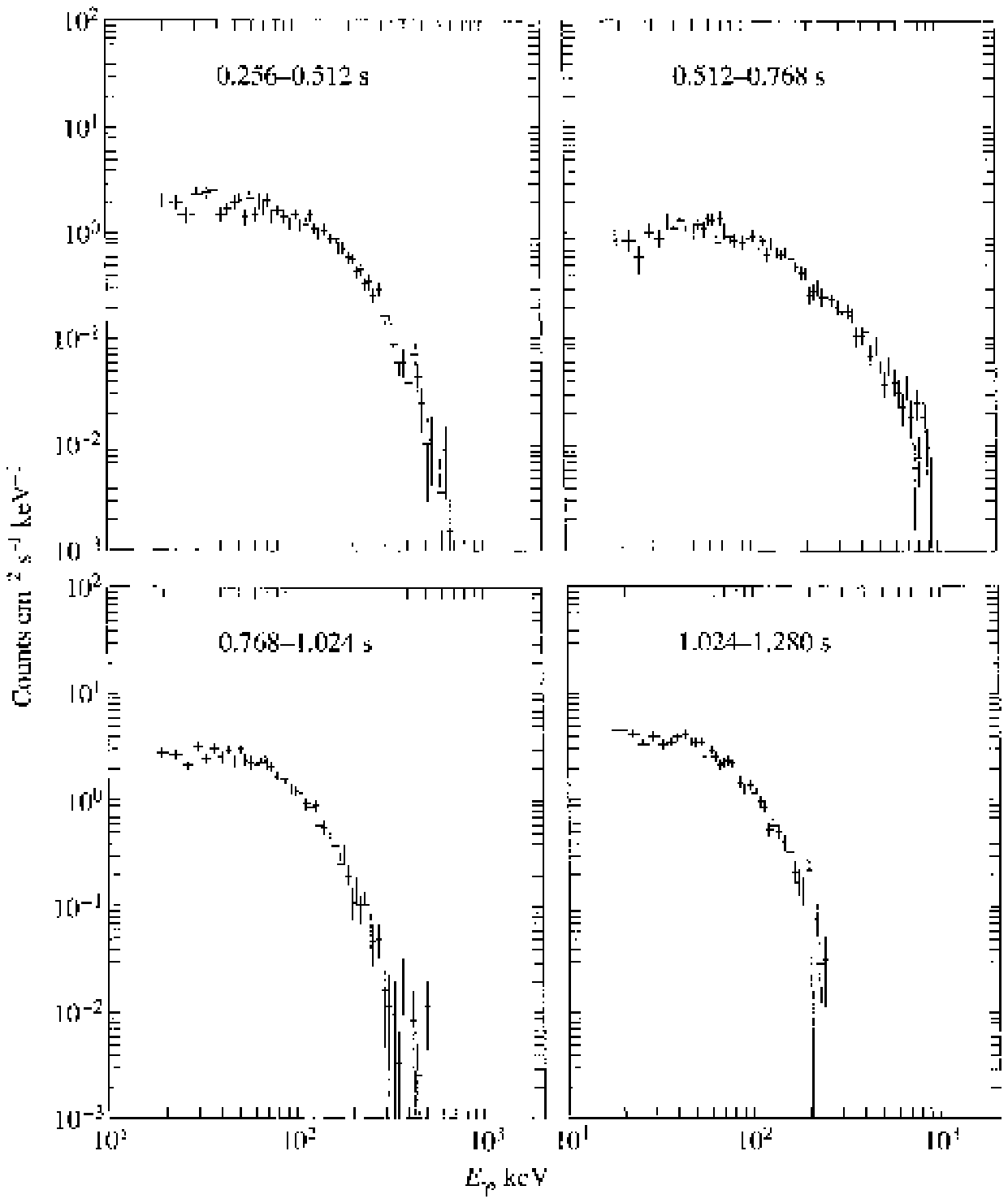}&
\includegraphics[width=6cm,height=7.5cm]{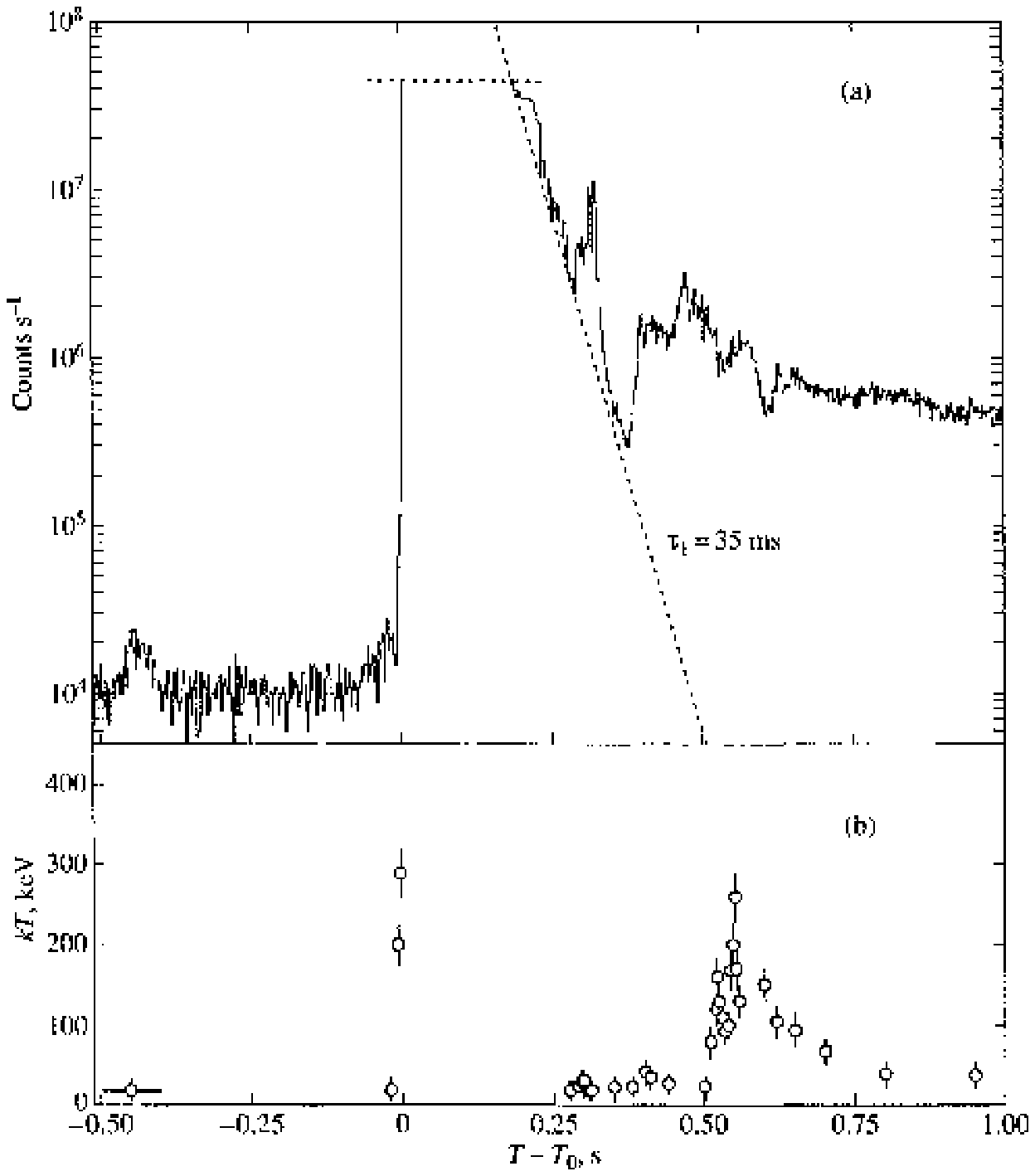}
\end{tabular}
\caption{\label{fig1} {\bf On the left:}
Spectra measured by Konus-Wind 
in 4 time intervals of the initial emission.
{\bf On the right:} Counting rate during the first second 
of the burst for $E_{\gamma} > 15$ keV and parameter kT of
the OTTB law $E^{-1} e^{-E/kT}$ that fits the energy spectra on the left
as a function of time.
Figures from \protect\cite{Mazets}.}
\end{figure}

Most information on the giant flare comes from {\it Beppo}-SAX 
\cite{Feroci,Guidorzi}, Konus-Wind \cite{Mazets} and
Ulysses \cite{Feroci1}. All suffer from saturation problems during the 
first second of the burst. Unfortunately, there are no clear 
spectral measurements covering the first $\sim 0.35$ s above 250~keV. 
Konus-Wind reported spectra below 250~keV for the first second of the 
emission; see Fig.~\ref{fig1} \cite{Mazets}. 
Also shown is the counting rate showing the saturation of the detector. 
Konus-Wind spectra can be described by a OTTB spectrum typical of 
electromagnetic processes because it was not sensitive to the 
high energy component.

The energy and fluence measurements are also complicated by pile-up 
and dead time problems. Mazets {\sl et al.} \cite{Mazets} find a lower 
limit to the fluence in the first $\sim 0.35$~s of $5.5 \cdot 10^{-3}$ 
ergs/cm$^2$ for $E_{\gamma} > 15$ keV. For a source at 14.5~kpc and isotropic 
emission this corresponds to a lower 
limit on the total energy emitted by the source of 
$1.4 \cdot 10^{44}$ ergs during the peak of the burst.
A similar amount of energy was emitted in the long tail with 
an exponential decay of $ \sim  90$ s that followed the burst and 
lasted several hundreds of seconds.

Similar conclusions can be inferred from the Ulysses data \cite{Feroci1}
that were even more affected by saturation because of its sensitivity 
in the interval 25-150~keV.

We next concentrate on the {\it Beppo}-SAX data with energy up to 
700~keV.

\subsection{{\it Beppo}-SAX observations of the Aug. 27, 1998 giant flare}
\label{sub2}

\begin{figure}[htb]
\begin{tabular}{cc}
\includegraphics[width=7cm,height=9cm]{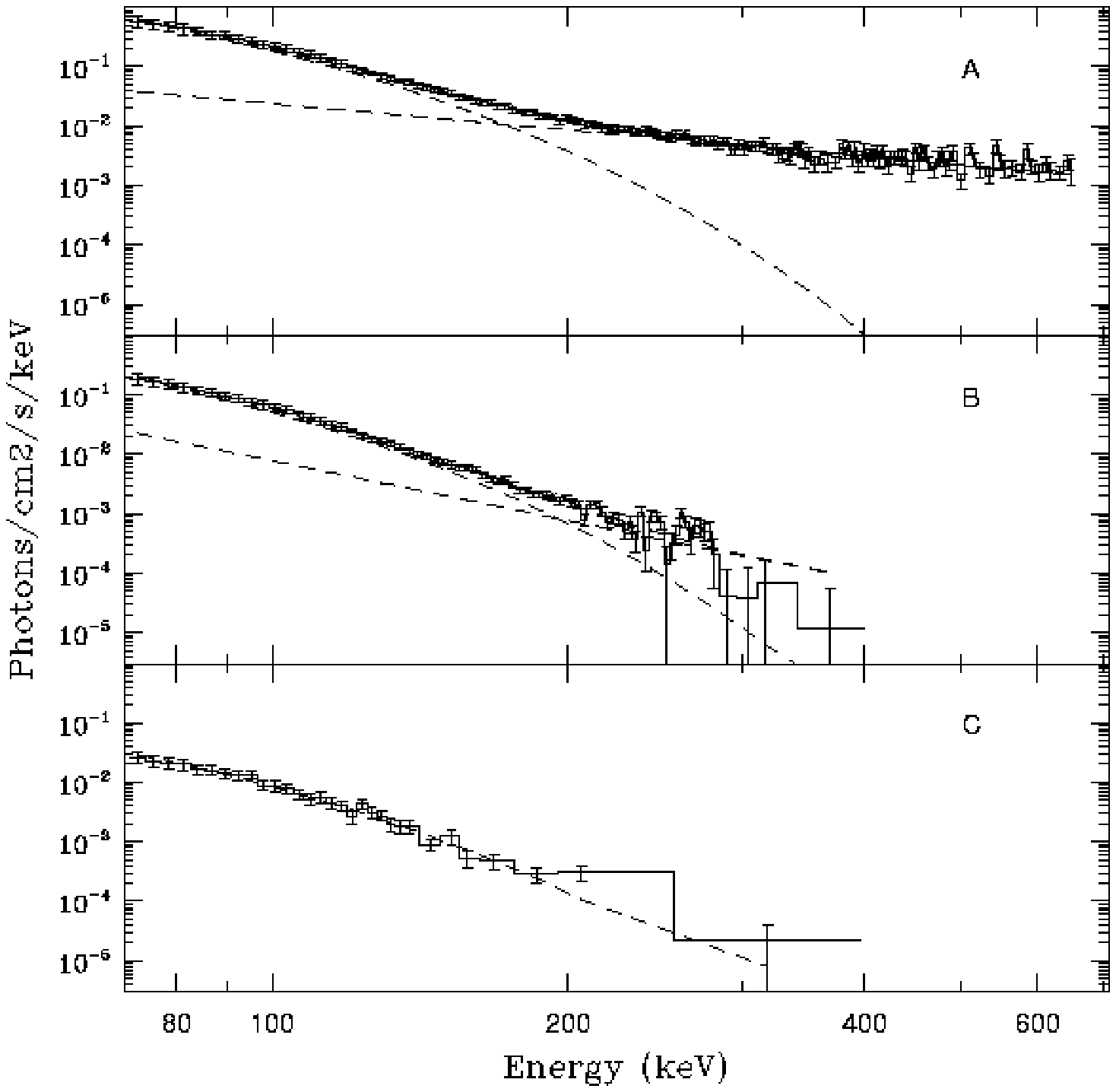}&
\includegraphics[width=6.cm,height=6cm]{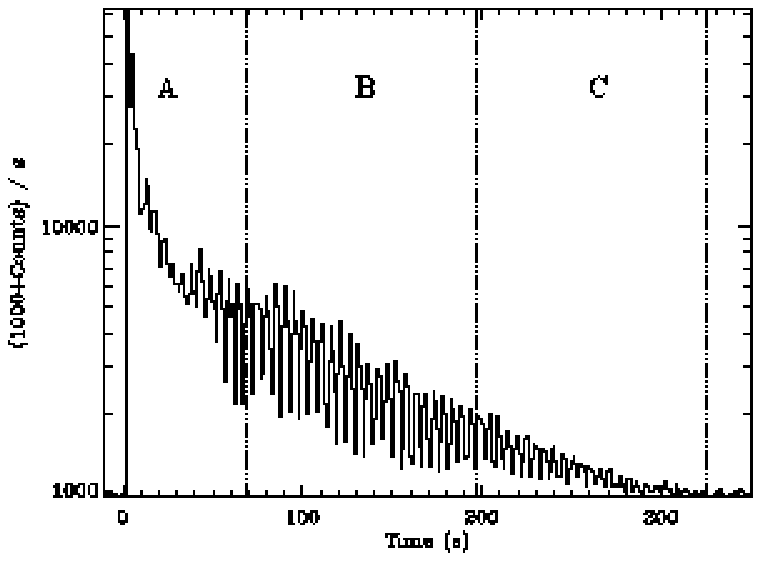}
\end{tabular}
\caption{\label{fig2} {\bf On the left:}
{\it Beppo}-SAX counting rate during the SGR 1900+14 flare: the 3 
intervals where the 
spectral analysis is done are 0-67~s (interval A), 68-195~s (interval B) and
196-323~s (interval C). {\bf On the right:}
The spectra for the 3 time intervals in the energy range of 40-700~keV. 
The best fit functions are shown too
(figures from \protect\cite{Feroci}).}
\end{figure}

{\it Beppo}-SAX data on SGR 1900+14 were first published in 
\cite{Feroci} and updated in \cite{Guidorzi}. Data from the 
{\it Beppo}-SAX Gamma-Ray Burst Monitor (GRBM) were also affected 
by saturation and only a lower limit on the fluence is reported for 
the first 68~s of the flare. The experiment could not determine the event rise-time 
for timescales shorter than 1 s. Fig.~\ref{fig2} shows the counting rate 
as well as the energy spectra corresponding to 3 different time intervals in the 
70-650~keV interval. Most importantly, the experiment provides information 
during the first 68 s where the data exhibit a very hard non-thermal 
component. The results were updated in \cite{Guidorzi} for a better 
understanding of the response of the GRBM. The energy range 
was extended to 40-700~keV.

\begin{figure}[htb]
\begin{tabular}{cc}
\includegraphics[width=6.5cm]{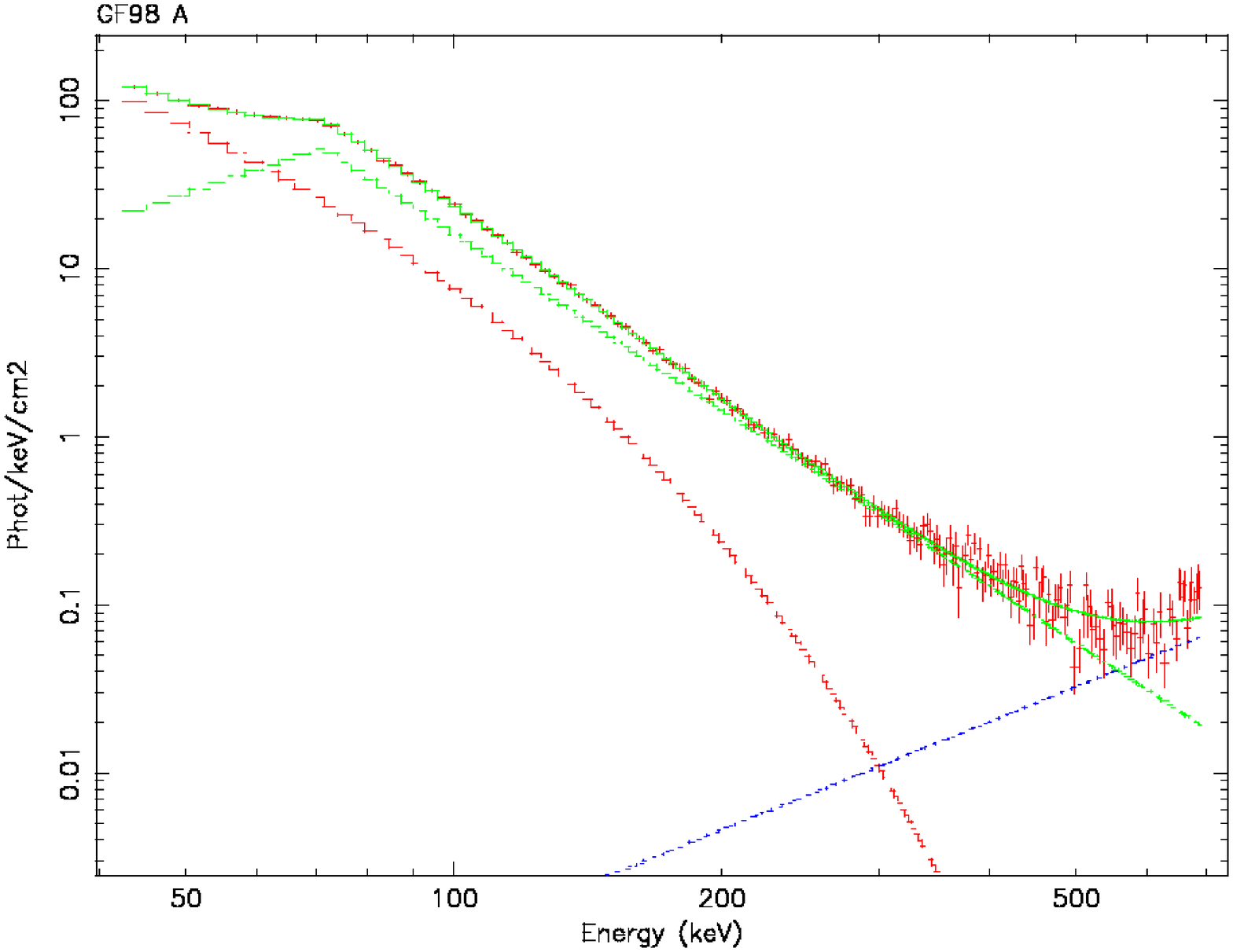}&
\includegraphics[width=6.5cm]{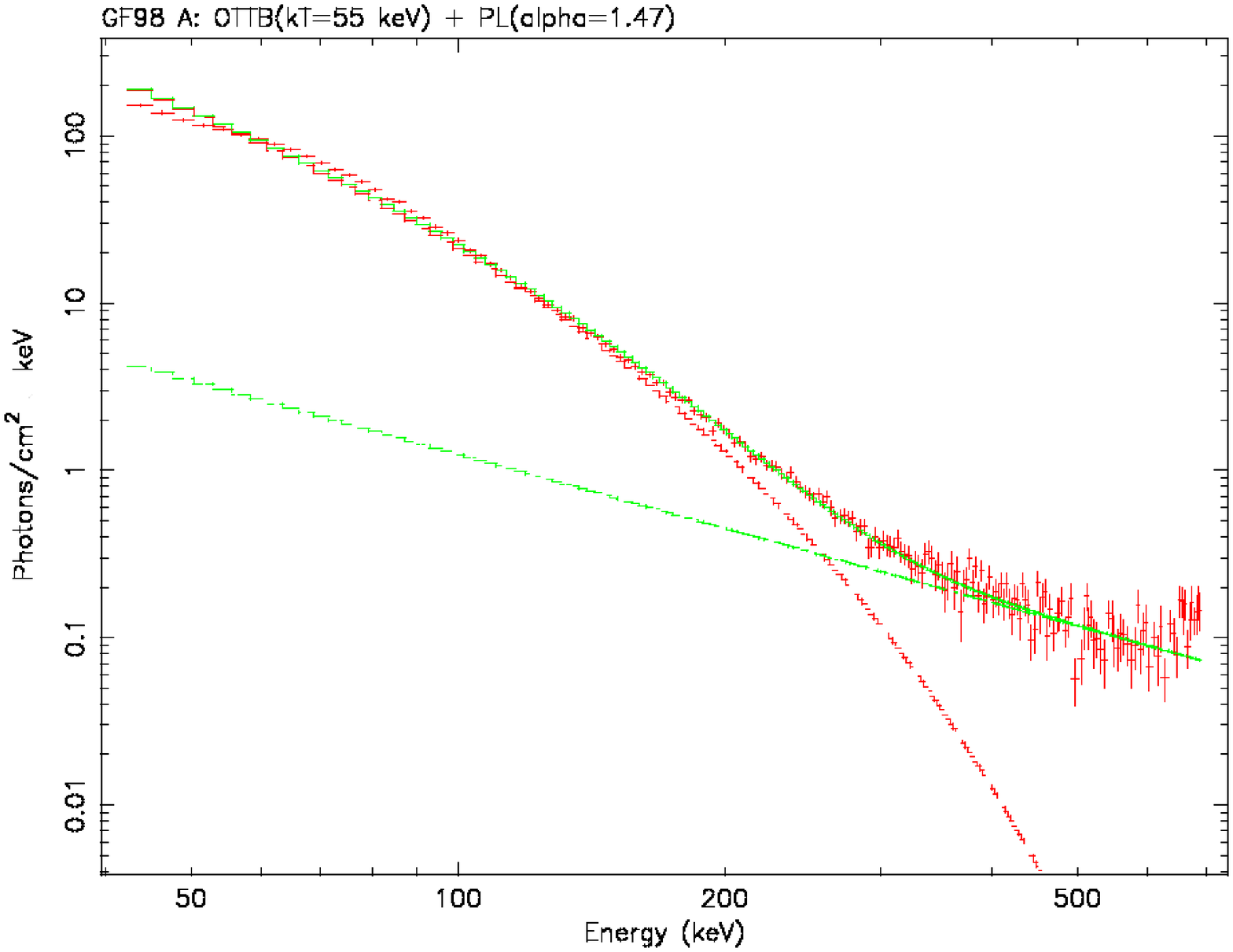}
\end{tabular}
\caption{\label{fig3} {\bf On the left:}
Best fit to the spectrum measured by {\it Beppo}-SAX during the first 68~s 
of the flare (referred to as GF98 A in the plots) 
in the 40-700~keV range.
{\bf On the right:} fit using the Feroci {\it et al.} OTTB+PL 
function \protect\cite{Feroci,Guidorzi1}.
The corresponding fit functions are given in Tab.~\ref{tab2}.
Courtesy of C.~Guidorzi \protect\cite{Guidorzi1}.}
\end{figure}

Possible fits to the data are presented in Fig.~\ref{fig3}. The published 
best fit is a sum of 3 functions given in Tab.~\ref{tab2}. 
We have instead chosen to adopt the previous fit in 
Feroci {\em et al.}, also shown in Fig.~\ref{fig3}, because it gives a 
better description of the high energy component to which we are
mostly interested in this work. The data are fit 
with OTTB + PL, while in the more recent paper a thermal bremsstrahlung 
spectrum (based on \cite{BREM})+ BKNPL + PL 
(with positive spectral index, probably an artifact of the large errors in 
the interval fitted) was adopted. The $\chi^2$ of the OTTB+PL is worse 
because the low energy part of the spectrum is poorly fitted by the 2 
functions compared to the BREM+BKNPL+PL fit. 
We also note that in the updated analysis the systematic error was 
reduced from 10\% to 2\% in \cite{Feroci} in the low energy range
and the minimum considered energy was decreased from 70~keV to 40~keV.

The total fluence corresponding to the 3 intervals is:
$>6.4$ for interval A \footnote{Notice that
this number is obtained as the integral of the functions 
in Tab.~\protect\ref{tab2}
multiplied by the duration of the burst $\Delta t = 68$ s.}, 
2.1 for interval B and 0.49 for interval C, in units of $10^{-4}$ erg/cm$^2$. 
Assuming an isotropic emission, the energy emitted by a source at 15.1 kpc is:
$>1.7$, 0.6, 0.1 in units of $10^{43}$ erg.
As already observed, the value in the first interval is smaller 
than the one reported by Konus-Wind for the first 0.35 s \cite{Mazets}.
 
\begin{figure}[htb]
\begin{tabular}{cc}
\includegraphics[width=7.cm]{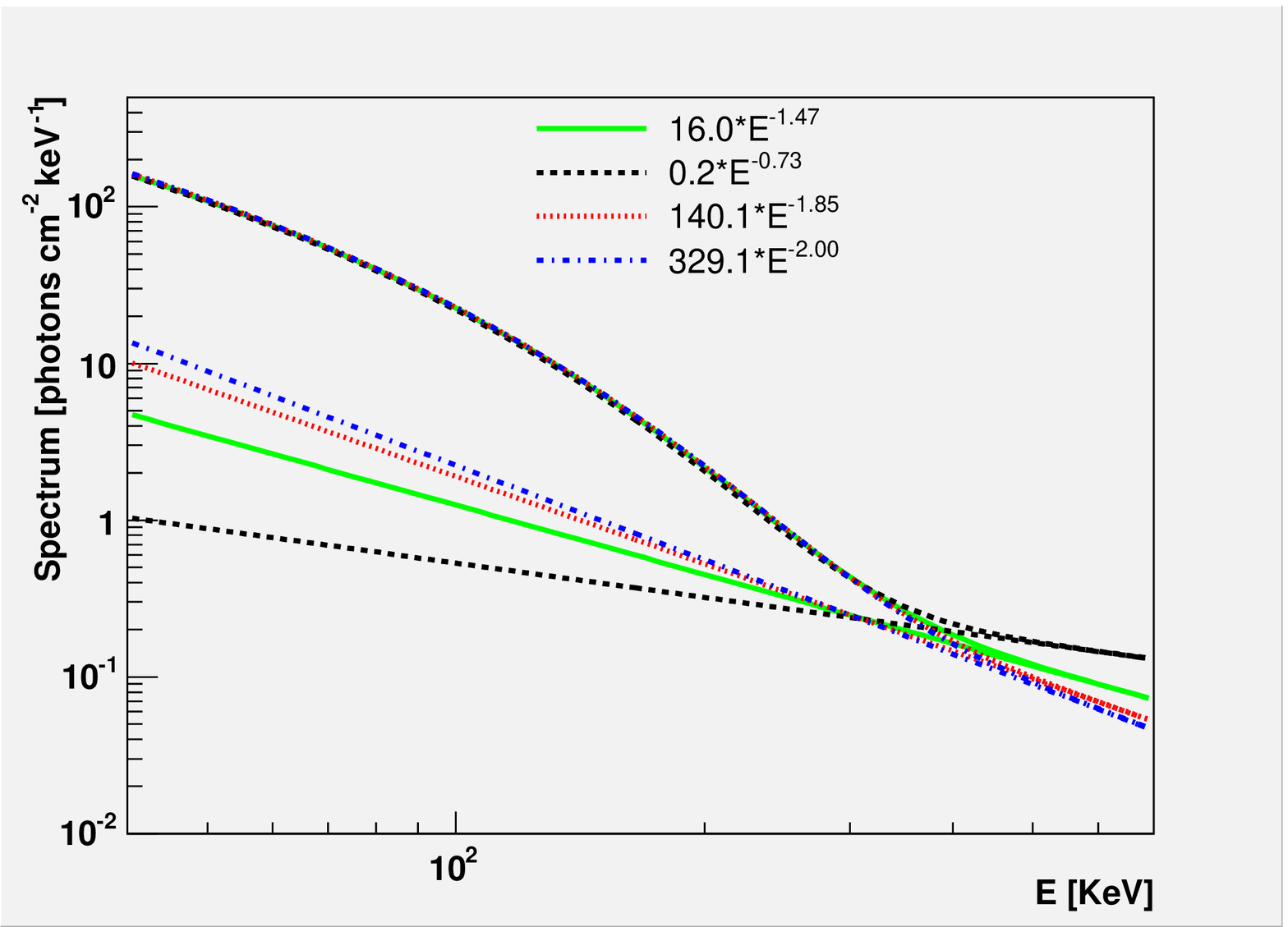}&
\includegraphics[width=7.cm]{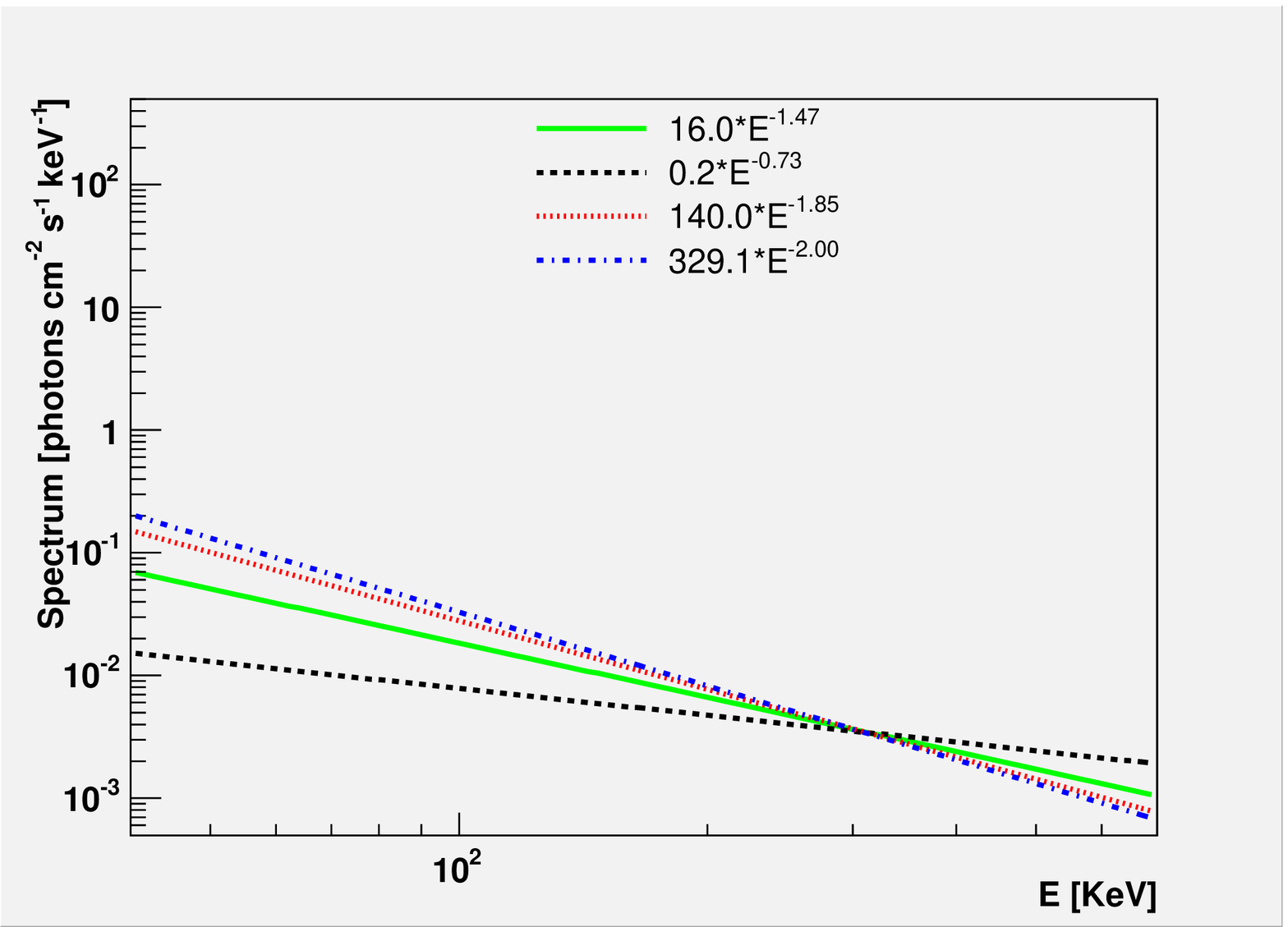}
\end{tabular}
\caption{\label{fig4} {\bf On the left:} 
Upper curve consists of the OTTB with kT = 55 keV + the 4 different PL components.  
Units are the same as in {\it Beppo}-SAX paper, keV$^{-1}$ cm$^{-2}$  
\protect\cite{Guidorzi}. This plot
shows that the different PL components do not alter much the total
spectrum at low energy. As a matter of fact all different PL's added to the OTTB
component overlap (see upper curves).
{\bf On the right:} 
Detail of the PL functions in the units used in this paper of 
keV$^{-1}$ cm$^{-2}$ s$^{-1}$ as
photon differential energy spectra for further estimates.
}
\end{figure}

For further calculations we will adopt a differential energy spectrum for 
$E>250$~keV of $\frac{dN}{dE} \propto E^{-1.47}$. We will study variations 
of the PL spectral index to allow for the uncertainties; see Fig.~\ref{fig4}. 
Moreover, since typical Fermi acceleration processes are characterized by 
an $E^{-2}$ spectrum, and given the suggestion that SGR 1806-20 may be a 
soft GRB \cite{Piran}, we also consider this assumption for the spectral 
shape. Our different assumptions are summarized in Tab.~\ref{tab3}. 
Also given is the fraction of the fluence in the PL component relative to 
the total fluence measured by {\it Beppo}-SAX. From Fig.~\ref{fig4} it is clear that 
the different functions adequately fit the data that were initially fitted 
as ($16 \cdot (E/{\rm keV})^{-1.47}$ keV$^{-1}$ cm$^{-2}$ s$^{-1}$). For each spectral
shape, the normalization was chosen so that the fluence 
in the interval 200-500 keV is the same as for the 
$E^{-1.47}$ fit in \cite{Feroci}.


\begin{table}[htb]
\begin{tabular*}{0.98\textwidth}{@{\extracolsep{\fill}} |c c c c c|} 
\hline
\hline
$1.1 \cdot 10^3 E^{-1}e^{-\frac{E}{40.3}}$ &+ &  $ 3.8 \cdot 10^{-4} E^{1.79}$ ($E<71 keV$) &+ &   $8.0 \cdot 10^{-10} E^{+2.14}$ \\
& &   $2.2 \cdot 10^6 E^{-3.48}$ ($E>71 keV$) &  &\\
\hline
BREMS & + & BKNPL & + & PL \\
\hline
\hline
\end{tabular*}
\begin{tabular*}{0.98\textwidth}{@{\extracolsep{\fill}} | c c c c c |}

&   $192.8 E^{-1} e^{-E/55}$ & + & $16.0 E^{-1.47}$ &  \\
\hline
&  OTTB                     & + & PL &  \\
\hline
\hline
\end{tabular*}
\caption{\label{tab2} 
Functions used by 
Guidorzi {\it et al.} (BREMS+BKNPL+PL) \protect\cite{Guidorzi,Guidorzi1}
and by Feroci {\it et al.} (OTTB+PL)
\protect\cite{Feroci,Guidorzi1}  
to fit the spectrum  for the first 68~s of the flare, respectively in the 
energy ranges 70-650~keV and 40-700~keV (see Fig.~\protect\ref{fig3}).
Energy is in keV.}
\end{table} 

\begin{table}[htb]
\begin{tabular}{|c|c|}
\hline
{\bf Energy Spectrum} & {\bf Fraction of measured fluence in 68 s} \\
{\bf (keV$^{-1}$ s$^{-1}$ cm$^{-2}$)}  &  {\bf } \\ \hline
$0.2 \times (E/keV)^{-0.73}$ & 12.9\%\\
${\bf 16.0 \times (E/keV)^{-1.47}}$ & {\bf 12.2\%}\\
$140.0 \times (E/keV)^{-1.85}$ & 14.8\%\\
$329.1 \times (E/keV)^{-2.00}$ & 16.0\%\\
\hline
\end{tabular}
\caption{\label{tab3} 
Differential photon energy spectra assumed in this paper.
The fraction of the total fluence $6.4 \cdot 10^{-4}$ ergs 
measured by {\it Beppo}-SAX 
between 40-700 keV that the PL components
would account for is also reported.
Numbers in bold are for the function used 
in \protect\cite{Feroci}.}
\end{table} 

\section{TeV-Gamma Showers and Induced Underground Muons}
\label{Sec3}

High energy gamma-rays initiate electromagnetic showers in the atmosphere
via the dominant processes of bremsstrahlung and pair production. 
With the development of the shower a large population of photons emerges 
that have a probability to produce a pion rather than an electron pair. 
Such pions decay into muons that, with sufficient energy, can penetrate to the 
depth of an underground detector. The fact that the process is rare is 
compensated by the large area of a detector such as AMANDA 
\cite{Yodh,Alvarez,Hooper}. The showers can also be directly observed 
by ground-based wide field of view gamma detectors such as Milagro \cite{Yodh}.

The high energy muons can be detected by deep under-ice/water
neutrino telescopes above the muon threshold determined by the detector depth,
geometry and reconstruction capabilities,
as observed in \cite{Yodh,Alvarez,Hooper}.
In order to derive the number of muons from a given photon spectrum
we follow the
analytical estimate in Ref.~\cite{Halzengamma}, that was checked
against the Monte Carlo calculation in Ref~\cite{Stanev}. We repeat here 
the main steps of that calculation.
The muon spectrum from a differential photon spectrum $\Gamma_{0}(E)$
is given by:
\begin{equation}
\frac{dN_{\mu}}{dE} = \Gamma_{0}(E) \frac{\epsilon_{\pi}}{E \cos\theta}
\left( \frac{1-r^3}{3(1-r)} \right) z_{\gamma\pi}\frac{\Lambda_{\pi}}{\lambda_{\gamma 
N}}\left[ 1+\ln\frac{t_{max}}{\Lambda_{\pi}}\right]
\label{eq:fluxmu}
\end{equation}
where energies are in TeV, $\theta$ is the zenith angle, 
$r = m_{\mu}^2/m_{\pi}^2$ is the ratio of the squared muon and pion masses, 
$\epsilon_{\pi} = 0.115$ TeV is the energy above which 
pion interaction dominates over decay.
The pion flux at the atmospheric depth t due to 
the photon flux $\Gamma_0$ can be factorized as
$\pi (E,t) = \Gamma_{0}(E) \pi_2 (t)$, with $\pi_{2}(t) = 
z_{\gamma\pi} \frac{\Lambda_{\pi}}{\lambda_{\gamma N}} (1 - e^{-t/\Lambda_{\pi}})$
$\sim z_{\gamma\pi} \frac{\Lambda_{\pi}}{\lambda_{\gamma N}}$. The various terms are
respectively: the z moment $z_{\gamma \pi} = \frac{1}{\sigma_{\gamma 
N}} \int_{0}^{1} dx x^{\alpha} \frac{d\sigma_{\gamma\rightarrow \pi}(x)}{dx}$,
that depends on the spectral index $\alpha$ of the photon flux $\frac{dN}{dE} \propto
E^{-(\alpha+1)}$;
the effective pion interaction length in the 
atmosphere $\Lambda_{\pi}$; the interaction length
associated with $\pi$ photo-production $\lambda_{\gamma N}$.   
Following the discussion in Ref.~\cite{Halzengamma} we assume:
\begin{equation}
z_{\gamma\pi} \frac{\Lambda_{\pi}}{\lambda_{\gamma N}} \sim
\frac{1}{1-z_{\pi\pi}} \sim \frac{A \sigma_{\gamma N}}{\sigma_{\pi 
A}} \times <nx>_{\gamma \rightarrow \pi} 
\end{equation}
with $z_{\gamma\pi}=\frac{2}{3}$ and $z_{\pi\pi}=\frac{2}{3}$ 
(this is the z moment for pion regeneration in the atmosphere). These values are 
suitable for hard spectra with $\alpha$ around 1. We assume 
a constant pion photo-production cross-section $\sigma_{\gamma N} \sim 0.1$~mb 
and an average mass number $A = 14.5$ in the atmosphere. This is a reasonable assumbtion
in the photon energy range $\sim 0.1-10^2$~TeV \cite{Battistoni,Stanev} 
\footnote{The resulting cross section from the
FLUKA transport and interaction code \protect\cite{FLUKA} is
in agreemet with this value of the $\gamma N$ cross section,
as well as data in Refs.\cite{FLUKAref} at GeV 
energies.}.
Moreover, we assume $\sigma_{\pi A} \sim 198$~mb \cite{Halzengamma}. 
The logarithmic term in eq.~\ref{eq:fluxmu}
depends on $t_{max}$, the depth
where the photon energy in the cascade has become too low
to produce muons of energy E.
It is assumed $t_{max} = \lambda_{R} ln\left[ 
\frac{E_{\gamma,max} <x_{\gamma\rightarrow\mu}>}{E} \right]$,
where $\lambda_{R}$ is the radiation length in the atmosphere.
$E/< x_{\gamma \rightarrow \mu} >$ is the $\gamma$ energy required to produce a
muon of energy E and we assume $< x_{\gamma \rightarrow \mu} >\sim 0.25$. Moreover, 
in eq.~\ref{eq:fluxmu} we have
$\frac{t_{max}}{\Lambda_{\pi}}=\frac{\lambda_{R}}{\Lambda_{\pi}} = (1-z_{\pi\pi})\frac{\sigma_{\pi 
A}}{\sigma_{R}}$ with $\sigma_{R} \sim 480$ mb.
The above muon flux is valid for energies large enough that 
muons do not decay, e.g. in our case, in which muons must have
enough energy to survive underground.

\begin{figure}[htb]
\begin{center}
\includegraphics[height=7.5cm,]{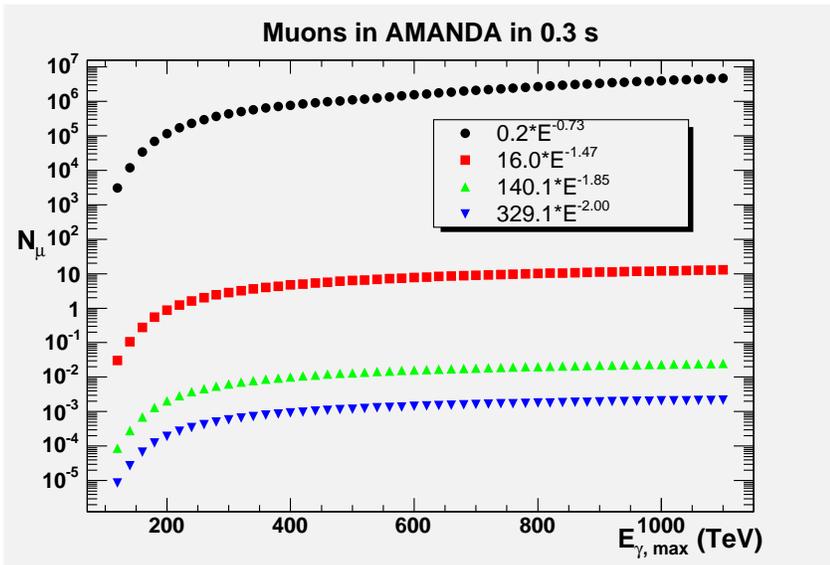}
\caption{\label{fig:mu} Number of muons induced by gamma showers in 0.3 s
for an almost horizontal area of 30000 m$^2$ for a detector
like AMANDA-II at a depth between 1.5-2 km for the 4 different photon
sectra indicated and as a function of the maximum photon energy. }
\end{center}
\end{figure}

 
The photon spectra motivated in Sec.~\ref{sub2} are expressed in the 
general form 
$\Gamma_0 (E) = \frac{dN}{dE} = \frac{F_{\gamma}}{E^{\alpha}} \cdot 10^{-12}$ 
with the units and parameters given in Tab.~\ref{tab3}. 
We assume that the spectra we get from SGR 1900+14 data 
are similar for SGR 1806-20, a burst with similar features and duration. 
Because SGR 1806-20 is more energetic by over one order of magnitude, 
this represents another conservative estimate. Moreover, we recall that 
the measured values are lower limits on the fluence for both bursts. 
We assume that the photon spectrum extends to 
$E_{\gamma, max}$ and in Tab.~\ref{tab4} we estimate the number 
of events for two values, 200 and 500~TeV, 
and show the number of signal events as a function
of $E_{\gamma,max}$ in Fig.~\ref{fig:mu}. Photons of energies
in excess of $\sim 500$~TeV should be attenuated over galactic distances
due to interaction on background photons.
The minimum muon energy $E_{\mu,min}$
that muons should have to cross the ice overburden depends on
the inclination of the source respect to the detector.
For AMANDA, located at the South Pole, a simple relation connects
the zenith angle and the declination of the source, that is
$\theta = \delta + 90^{\circ} = 70^{\circ}$. Being the detector 
at a depth of $1500-2000$ m, a muon induced by gammas from the source 
will have to transverse about 5.1 km of ice overburden before reaching the 
detector.
We calculate the survival probability
of muons with energy larger than 10 GeV at AMANDA depths 
using the MUM code for high energy muon energy losses \cite{MUM}.
In this calculation we account for the muon spectrum and the parent gamma
spectrum in Tab.~\ref{tab3}.
Hence we derive the number of muons at AMANDA depth 
and for the given source zenith as:
\begin{equation}
N_{\mu} = \int_{E_{\mu,min}}^{E_{\mu,max}} \frac{dN_{\mu}}{dE} 
P_{surv}(depth,E_{\mu}) dE_{\mu}  
\label{eq:mu}
\end{equation}
assuming that $E_{\mu,max} \sim <x_{\gamma\rightarrow\mu}> 
E_{gamma,max} \sim 0.25 E_{\gamma,max}$, for depth equal to 5.1 km.

The main background source to the muons produced by gammas emitted in the flare
are atmospheric muons. In the horizontal direction the atmospheric muon rate
is reduced by roughly 2 orders of magnitude with respect to 
the vertical direction. Therefore the background is about
1 Hz. Furthermore, since the bulk of the energy is emitted in only 0.3 s from a 
point-like source, 
the background is limited to the time interval and the direction of the burst. 
Conservatively considering a search bin of 8$^{\circ}$, well
above the resolution of the detector and comparable to the
size of the declination band \cite{AMANDApoint} chosen by the
experiment after a signal to noise optimization,
the estimated background rate from atmospheric muons is at the level 
of $\pi (8*\pi/180)^2/(2\pi) \times 0.3 \times 1 = 3 \cdot 10^{-3}$ events
in 0.3 s. Hence it is negligible.

Where the signal is concerned, we tabulate in Tab.~\ref{tab4} the expected
number of muons from the burst in 0.3 s for $E_{\gamma,max} = 200$ TeV
and 500~TeV. 
We assumed a near horizon effective area 
close to the geometrical area of AMANDA of about 30000 m$^2$. The effective 
area includes cuts on the quality of the muon reconstruction, and was estimated for
harder muon spectra than atmospheric muons for which an average value is 22000 m$^2$.
In Fig.~\ref{fig:mu} we show how the results vary for different $E_{\gamma,max}$ values
and for the 4 spectra
we considered. Similar event rates might be observed by the future ANTARES detector, located
at $2050-2400$ m in the sea, that would be able to 
observe muons and showers induced by neutrinos from another flare from the source, 
SGR 1900+14, for more than 40\% of a day. It will also be able to observe 
neutrino induced upward going muons and showers from SGR 1806-20. 

As expected, the event rate is a strong function of the
spectral index and of the normalization. In the absence of a possible detection, 
it is straightforward for AMANDA to constrain the high energy flux associated 
with the possibility that SGR's are high energy accelerators, 
possibly similar to GRB's.

\begin{table}[htb]
\begin{tabular}{|c|c|c|c|c|}
\hline
{${\bf F_{\gamma}}$} & {\bf Spectral} & {\bf $\mu$ Events} & {\bf 
Atmospheric  }&{\bf $\gamma$
showers} \\ 
{\bf } & {\bf Index} & {\bf  AMANDA} & {\bf $\mu$'s in AMANDA}& {\bf in Milagro}\\ \hline
$6.47 \cdot 10^{13}$ & -0.73 & $0.1-1 \cdot 10^6$ & $3 \cdot 10^{-3}$ & 
$1 \cdot 10^{11}$\\ 
{$\bf 9.42 \cdot 10^8$}& {\bf -1.47} & {\bf 0.8-6}& ${\bf 3 \cdot 10^{-3}}$&
${\bf 8 \cdot 10^{4}}$\\
$3.13 \cdot 10^{6}$& -1.85& $0.1-1 \cdot 10^{-2}$ &$4 \cdot 10^{-3}$&79\\ 
$3.29 \cdot 10^{5}$& -2.00& $0.1-1 \cdot 10^{-3}$&$2 \cdot 10^{-4}$& 5\\
\hline
\end{tabular}
\caption{\label{tab4} The normalization of the
$\gamma$ flux and the spectral index for the giant flare.
The normalization $F_{\gamma}$ units are such that the
differential flux 
$\frac{dN}{dE} = F_{\gamma}\cdot (E/TeV)^{\alpha} \cdot 10^{-12}$
is in cm$^{-2}$ s$^{-1}$ TeV$^{-1}$.
The values of the {\it Beppo}-SAX fit are in bold.
The corresponding
number of muons in a detector like AMANDA-II of 30000 m$^2$ in 0.3 s is given.
Two numbers are given in the third column: the smaller is obtained assuming
a maximum photon energy of 200~TeV and the larger assuming 500~TeV. 
Also the atmospheric muon background in a search solid angular bin of 
$8^{\circ}$ and in 0.3~s is indicated. 
The gamma shower events in Milagro in 0.3 s are also shown for a giant flare in
its field of view (see the text).}
\end{table} 

\subsection{Estimate of photon induced showers in Milagro}
\label{sub3}

SGR 1806-20 was located above the Milagro horizon at the time of the Dec. 
27 giant flare at a zenith angle of about $68^{\circ}$. 
The source
might have been visible for Milagro though we understand data at such zenith
angle are not routinely analyzed by the experiment whose range covers 
$0-60$ degrees.
Nonetheless, we consider the effective area presented in 
\cite{Atkins,Hooper} as function of the $\gamma$ energy:
$A(E_{\gamma})=4 \cdot 10^6 (E/TeV)^{\beta}$
in cm$^2$, where $\beta = 1.39$ for $E_{\gamma} > 1$~TeV and
$\beta=2.35$ for $E_{\gamma} \in [0.3,1]$~TeV,
and we use it for estimating the number of shower events
in Milagro from an SGR burst eventually happening in a favorable
sky region.
Using the various gamma spectra we estimate event rates in Tab.~\ref{tab4}
using the following formula:
\begin{equation}
\frac{N_{showers}}{T} = \int_{E_{\gamma,min = 0.3 TeV}}^{E_{\gamma,max}=100 TeV}\frac{dN_{\gamma}}{dE_{\gamma}} \times A(E_{\gamma}) dE_{\gamma}\, .
\label{eq:showers}
\end{equation}
For some of the spectral indexes Milagro may have seen a large signal
during the first instants of the flare.

\section{Neutrino initiated showers and upward-going muons from giant SGR 
flares}
\label{sec:neu}

The photon fluxes previously discussed are accompanied by neutrinos 
because charged pions are produced at the source along with the neutral pions that decay 
into photons. Assuming that photons and neutrinos have this common origin, 
the neutrino spectrum can be derived from the photon spectrum by energy 
conservation along with simple kinematics \cite{Alvarez,Hooper1}. 
Photons may cascade in the source to emerge with a steeper spectrum and 
they may even be absorbed, therefore only a lower limit on the neutrino 
flux can be derived using this technique.

It is likely that the target for neutrino production are the thermal photons 
surrounding the pulsar. We therefore assume $p-\gamma$ interactions with 
the ambient radiation around the neutron star. There may also be a 
contribution from neutron decay from photo-disintegration of nuclei.
The relation between the two spectra is:

\begin{equation}
\int_{E_{\gamma,min}}^{E^{\gamma,max}} E_{\gamma} \frac{dN_{\gamma}}
{dE_{\gamma}} dE_{\gamma} = K \int_{E_{\nu,min}}^{E_{\nu,max}} 
E_{\nu} \frac{dN_{\nu}}{dE_{\nu}} dE_{\nu} \, ,
\label{eq:1}
\end{equation}
where K = 1 for $pp$ interactions and 4 for $p-\gamma$. 
In $pp$ interactions charged and neutral pions are produced in equal 
numbers yielding 2 $\gamma$'s carrying 1/2 of 
the neutral pion energy, that is 1/3 of the proton energy, and so 
$E_{\gamma} \sim E_{p}/6$. If muons also decay, there are 2 muon
neutrinos per $\gamma$ and so $E_{\nu} \sim E_{p}/12$.
For $p-\gamma$ interactions, neutrinos and photons are produced from the 
$\Delta$ resonance decay: 
$p + \gamma \rightarrow \Delta \rightarrow
n \pi^{+} \rightarrow n \mu^{+} \nu_{\mu} \rightarrow n e^{+} \nu_{e} 
\bar{\nu}_{\mu} \nu_{\mu}$ and
$p + \gamma \rightarrow \Delta \rightarrow
p \pi^{0} \rightarrow p \gamma \gamma$.
Hence $E_{\gamma,min}$ is determined by the threshold of the 
process for protons \cite{Hooper1}:
\begin{equation}
E_{p} = \Gamma \frac{(2m_{p}+m_{\pi})^2-2m_{p}^{2}}{2m_{p}} \sim \Gamma \times 1.23 GeV 
\end{equation}
where we will consider a Lorenz factor $\Gamma \sim 1$.
The minimum neutrino and photon energy are about 5\% and 10\%, respectively, 
of the proton energy ($E_{\nu,min}$ = $<x_{p\rightarrow\gamma}>$ $E_{p,min}/4$ 
and $E_{\gamma,min} = <x_{p\rightarrow\gamma}>E_{p,min}/2)$, where 
$<x_{p\rightarrow\gamma}> \sim 0.2$ is the average fraction of proton energy
transferred to pions). 
The maximum proton energy, and hence the maximum gamma and neutrino 
energies, depend on the kinematics in the beam that results from the
burst.
We will assume values a bit larger than what is calculated in 
Ref.~\cite{Zhang}:  here $E_{p,max} \sim 100-300$~TeV \cite{Zhang}, where 
protons are accelerated in the potential drop of the magnetar with surface 
magnetic field of $\sim 10^{15}$ Gauss. As a matter of fact, we are
considering giant flares and not the steady emission.
Hence, we will assume here a maximum photon energy of 200~TeV and a
maximum neutrino energy of 100~TeV. Also we assume that the photon 
spectra have the same spectral index, 
a very conservative assumption \cite{Hooper}.
Using the $\gamma$ spectra in Tab.~\ref{tab3}, we obtain the neutrino fluxes 
in Tab.~\ref{tab5}.

\begin{table}[htb]
\begin{tabular}{|c|c|c|c|c|}
\hline
{\bf Neutrino Spectrum} &  \multicolumn{3}{|c|}{\bf Cascade events (s$^{-1}$)} & 
{\bf Upward-going $\mu$'s }  \\ 
{\bf (cm$^{-2}$ s$^{-1}$ GeV$^{-1}$)} & \multicolumn{1}{c}{\bf $\nu_e$}&
\multicolumn{1}{c}{\bf $\nu_{\mu}$}&\multicolumn{1}{c|}{\bf $\nu_{\tau}$} 
& {\bf ($s^{-1}$)}  \\ \hline
5.90(E/GeV)$^{-0.73}$ & $1 \cdot 10^{7}$ & $8 \cdot 10^{6}$ 
& $1 \cdot 10^{7}$ & $2 \cdot 10^{6}$\\
{\bf $8.74 \cdot 10^{-3}(E/GeV)^{-1.47}$} & {\bf 0.4}& {\bf 0.2}&
{\bf 0.3} & {\bf 0.8}\\
$3.09 \cdot 10^{-4}(E/GeV)^{-1.85}$& $9 \cdot 10^{-5}$ & $3 \cdot 10^{-5}$ 
& $6 \cdot 10^{-5}$ & $5 \cdot 10^{-4}$\\ 
$8.23 \cdot 10^{-5}E^{-2.00}$& $3 \cdot 10^{-6}$ & $1 \cdot 10^{-6}$ &
$2 \cdot 10^{-6}$& $3 \cdot 10^{-5}$\\
\hline
\end{tabular}
\caption{\label{tab5} Neutrino spectra obtained using the
fluxes in Tab.~\protect\ref{tab3} and eq.~\protect\ref{eq:1} and shower and
upgoing-muon event rates in a detector like
AMANDA-II per second for a source at $\delta = \pm 20^{\circ}$.
}
\end{table} 

We use these fluxes to derive event rates for two different neutrino 
signatures: neutrino induced showers (cascade events), that have a 
$4\pi$ coverage and can be observed even if
the source is in the upper hemisphere of the detector and
neutrino induced upward-going muons that can be observed only if the 
source is in the lower hemisphere; this is the case for SGR 1806-20 
in the field of view of 
ANTARES and for SGR 1900+14 observable by AMANDA-II.
We estimate event rates for AMANDA neutrino effective areas, 
though similar numbers apply for the present ANTARES detector design.
Neutrino effective areas as a function of the energy for 
the declination of a source can be directly convoluted
with the neutrino spectra to provide event rates.
To estimate the number of cascade events we have considered the
AMANDA effective areas for neutrinos of all flavors
presented in  \cite{AMANDAcascades}. Because of oscillations, a 
cosmic neutrino beam reaches the Earth with equal fluxes for all flavors. 
For the near horizontal direction of the sources under consideration 
the absorption of neutrinos is almost negligible. 
To estimate upward-going neutrino event rates
we used the effective area presented in \cite{AMANDApoint}.
Results are given in Tab.~\ref{tab5}.

The background for neutrinos from the flare is due to atmospheric neutrinos and
muons. As already explained in Sec.~\ref{Sec3}, it can be
rejected using time and directional constraints. We neglect the
contribution to the neutrino signal from the flare
due to neutrinos that accompany muons from gammas considered in
Sec.~\ref{Sec3} because they would give a much smaller contribution to
neutrinos directly produced at the source.

\section{Conclusions} 

Starting from measured spectra by {\it Beppo}-SAX for the SGR1900+14
giant flare, we have estimated the TeV gamma-ray and neutrino event rates 
expected (and possibly observed) for the AMANDA detector from SGR 1806-20 
on Dec. 27, 2004. We showed that tight time and angular constraints
can reduce the atmospheric muon background to negligible levels.
The detectors are essentially free of background because of the time 
constraints.

A negative result will result in an interesting constraint on the
possibility that SGR most intense flares are similar to small 
GRB's in our Galaxy.

\section*{Acknowledgments}

We thank Cristiano Guidorzi for
helping us in understanding {\it Beppo}-SAX measurement
and Igor Sokalski for providing us with underground muon survival
probabilities.
We also would like to thank the UW IceCupe group, and particularly
Paolo Desiati and Albrecht Karle, for
fruitful discussions.

\end{document}